\documentclass[aps,pra,twocolumn,superscriptaddress,floatfix]{revtex4-1}
\usepackage{graphicx}
\usepackage{amssymb,amsmath}

\def\th         {\ensuremath{^\mathrm{th}}}
\def\st         {\ensuremath{^\mathrm{st}}}

\def\qed        {\mbox{ }~\hfill~$\Box$}

\newcommand{\ket}     [1] {\ensuremath{\left|#1\right\rangle}}
\newcommand{\Tone}{\ensuremath{\mathrm{T_1}}}

\def\eps  {\varepsilon}
\def\up   {\uparrow}
\def\down {\downarrow}

\def\FF   {\ensuremath{\mathcal{F}}}
\def\BB   {\ensuremath{\mathcal{B}}}

\begin{document}
\title{Semi-optimal Practicable Algorithmic Cooling}
\author{Yuval Elias} 
\affiliation{Schulich Faculty of Chemistry, Technion, Haifa, Israel}
\author{Tal Mor} 
\affiliation{Computer Science Department, Technion, Haifa, Israel}
\author{Yossi Weinstein}
\affiliation{Physics Department, Technion, Haifa, Israel}
\date{March 1 2011}

\begin{abstract}
\emph{Algorithmic Cooling (AC) of spins} applies entropy manipulation
algorithms in open spin-systems in order to cool spins far beyond Shannon's
 entropy bound. AC of nuclear spins was demonstrated experimentally,
 and may contribute to nuclear magnetic resonance (NMR) spectroscopy.
Several cooling algorithms were suggested in recent years, including 
practicable algorithmic cooling (PAC) and exhaustive AC\@. 
{\em Practicable} algorithms have simple implementations, yet their level of
cooling is far from optimal; {\em Exhaustive} algorithms, on the other hand,
cool much better, and some even reach (asymptotically) an optimal level of
cooling, but they are not practicable.

We introduce here {\em semi-optimal} practicable AC (SOPAC), wherein few 
cycles (typically 2--6) are performed at each recursive level.
Two classes of SOPAC algorithms are proposed and analyzed. Both attain cooling
 levels significantly better than PAC, and are much more efficient than the
 exhaustive algorithms. The new
algorithms are shown to bridge the gap between PAC and exhaustive AC\@.
In addition, we calculated the number of spins required by SOPAC in order to
purify qubits for quantum computation. As few as 12 and 7 spins are required (in an
ideal scenario) to yield a mildly pure spin (60\% polarized) from initial
polarizations of
 1\% and 10\%, respectively. In the latter case, about five more spins are sufficient
 to produce a highly pure spin (99.99\% polarized), which could be relevant for
fault-tolerant quantum computing.
\end{abstract}

\maketitle

\section{Introduction}

%
We consider here spin-half nuclei, quantum bits (qubits), whose
 computation-basis states align in a
 magnetic field either in the direction of the field ($\ket{\up} \equiv
 \ket{0}$), or in the opposite direction ($\ket{\down} \equiv \ket{1}$). 
Several such bits (of a single
molecule) represent a binary string, or a register. A macroscopic number of
such registers/molecules can be manipulated in parallel, as done, for instance,
in \emph{nuclear magnetic resonance (NMR)}.
 From the perspective of \emph{NMR quantum computation (NMRQC)}~\cite{SM08} (for
a recent review see~\cite{Jones10}), the
spectrometer that monitors and manipulates these qubits/spins 
can be considered a simple ``quantum computing'' device (see for instance Refs.
~\cite{BMRV10} and~\cite{NMR+06} and references therein).
The operations (gates, measurements) are performed in parallel on many registers.

%
The probabilities of a spin to be up or down are 
$p_{\up}=\frac{1+\eps_0}{2}$ and $p_{\down}=\frac{1-\eps_0}{2}$,  
where $\eps_0$ is called the \emph{polarization bias}
\footnote{$\eps_0\triangleq \tanh \left(\frac{\gamma\hbar B_0}{2K_B T}\right)$, 
where $K_B$ is the Boltzmann constant, $\gamma$ is the gyromagnetic coefficient,
$B_0$ is the intensity of the magnetic field and $T$ is the temperature of the 
heat bath. For small values, 
$\eps_0 \approx \frac{\gamma\hbar B_0}{2K_B T} \ll 1.$}. 
At equilibrium at room temperature, nuclear-spin polarization biases are 
very small; In common NMR spectrometers $\eps_0$ is at most 
$\sim 10^{-5}$. While the polarization bias may be increased by physical cooling
of the environment, this approach is very limited for liquid-state NMR,
especially for in-vivo spectroscopy.
To improve polarization, including selective enhancement, various ``effective
cooling'' methods increase the polarization of one or
more spins in a transient manner; The cooled spins re-heat
 to their equilibrium polarization as a result of thermalization with the 
environment, a process which has a characteristic time of \Tone.

%
An interesting effective cooling method is the reversible (or unitary) entropy 
manipulation of S{\o}rensen and Schulman-Vazirani~\cite{Sorensen89,SV99}, 
in which data compression tools and algorithms are used to compress entropy 
onto some spins in order to cool others. Such methods are limited by Shannon's 
entropy bound, stating that the total entropy of a closed system cannot 
decrease. The reversible cooling of Schulman-Vazirani was suggested as a new 
method to present scalable NMRQC devices. The same is true for 
the algorithmic cooling approach~\cite{BMRVV02} that is considered here. 

%
\emph{Algorithmic Cooling (AC)} combines the reversible effective-cooling
described above with thermalization of spins at different rates (\Tone\
values)~\cite{BMRVV02}. AC employs slow-thermalizing spins
(\emph{computation spins}) and rapidly thermalizing spins (\emph{reset spins}).
Alternation of data compression
 steps that put high entropy (heat) on reset spins, with thermalization of hot
 reset spins (to remove their excess entropy) can reduce the total entropy of
suitable spin systems far beyond Shannon's bound. Let us describe in detail the
three basic operations of AC\@. 
\emph{
 \begin{enumerate}
  \item \textbf{RPC\@. Reversible Polarization Compression} steps redistribute
 the entropy in the system so that some computation spins are cooled while other
   computation spins become hotter.
  \item \textbf{SWAP / POLARIZATION TRANSFER\@.} 
Controlled reversible interactions allow the hotter computation spins to 
   adiabatically lose their entropy to a set of designated reset spins.
Alternatively, we say that the higher polarization of the reset spins is
transferred onto these hotter computation spins.
  \item \textbf{WAIT / RESET\@.} 
The reset spins rapidly thermalize (much faster than the computation spins),
conveying their excess entropy to the environment, while the computation spins
remain colder, so that the entire system is cooled.
 \end{enumerate}
}
\noindent In some algorithms, the reset spins can be used directly in the
compression step~\cite{FLMR04, SMW05}. These algorithms have already led to 
 some experimental suggestions and implementations~\cite{FMW05, AAC-pat, POTENT,
 BMR+05, EFMW06, RMB+08}. 

%
An efficient and experimentally feasible AC technique, termed
 \emph{``practicable algorithmic cooling (PAC)''}~\cite{FLMR04}, 
combines \emph{polarization transfer (PT)}, reset, and
RPC among three spins called \emph{3-bit-compression (3B-Comp)}. 
The subroutine {\em 3B-Comp} may be implemented in several ways. 
One implementation is as follows~\cite{FLMR04}:
\\
\emph{
Exchange the states $\ket{100}\leftrightarrow\ket{011}$.\\ 
Leave the other six states ($\ket{000}$, $\ket{001}$, etc.) unchanged.
}\\
If 3B-Comp is applied to three spins \{C,B,A\} with biases $\eps_0 \ll 1$, then
 spin $C$ will acquire the bias~\cite{FLMR04}: $\eps' = \frac{3\eps_0}{2}.$
PAC1 and PAC2 consider \emph{purification levels}, where spins are cooled by a
 factor of $1.5$ at each successive level; At the $(j+1)\st$ purification level,
 $\eps_{j+1} = 1.5\eps_j = 1.5^{j+1}\eps_0$~\footnote{We are mostly concerned
 (with the exception of section~\ref{sec:pure-qubits}) with biases that are $\eps \ll
 1$, therefore calculations are usually done to leading order in $\eps$.}.

%
The practicable algorithm PAC2~\cite{FLMR04} uses any odd number of spins,
$n=2J+1$: one reset spin and $2J$ computation spins (see
Appendix~\ref{app:alg-PAC} for a formal description of PAC2). 
PAC2 cools the spins such that the coldest spin can,
ideally~\footnote{All calculations done here are ideal, in the sense that
the reset spins reset infinitely faster than the computation spins, and the
gates implementing the algorithms are error-free.}, attain a bias of $(3/2)^J$.
When all spins are cooled, the following biases are reached:
$\left\{\ldots,5\frac{1}{16},
3\frac{3}{8},
3\frac{3}{8},
2\frac{1}{4},
2\frac{1}{4},
1\frac{1}{2},
1\frac{1}{2},
1,1\right\}.$
This proves an exponential advantage over the best possible reversible cooling
techniques (e.g., of Refs.~\cite{Sorensen89} and~\cite{SV99}), which are limited
to a cooling factor of $\sqrt{n}$. As PAC2 can be applied to as few as
3 spins ($J=1$) to 9 spins ($J=4$), using NMRQC tools,
it is potentially suitable for near future applications~\cite{AAC-pat}.
PAC is simple, as all compressions are applied to three spins
and use the same subroutine, \emph{3B-Comp}. 

%
PAC2 always applies compression steps ({\em 3B-Comp}) to three identical biases.
In general, for three spins with biases, $(\eps_C,\eps_B,\eps_A)$, where all
biases $\ll 1$, 3-bit compression would attribute spin $C$ with the bias  
~\cite{EFMW06,Jose-PhD-Thesis,Thesis}: 
\begin{equation}\label{eq:3B-Comp}
\eps'_C = \frac{\eps_C+\eps_B+\eps_A}{2}\ .
\end{equation}
Using Eq.~\ref{eq:3B-Comp}, several algorithms were designed, including PAC3,
which always applies 3B-Comp to bias configurations of the form
 $\{\eps_j,\eps_j, \eps_{j-1}\}$ (see formal description in
 Appendix~\ref{app:alg-PAC}).
When all spins are cooled, the following biases are obtained:
$\left\{\ldots,9\frac{9}{16},
7,5\frac{1}{8},3\frac{3}{4},2\frac{3}{4},2,1\frac{1}{2},1,1\right\}.$

%
The framework of PAC was extended to include multiple cycles at each 
recursive level. Consider the simplest case, whereby a three-spin system
 $(C,B,A),$ with equal bias $\eps_0$, is cooled by repeating the following
 procedure $m$ times: 
\begin{enumerate}
\item {\em 3B-Comp} on $(C,B,A),$ increasing the bias of $C$.
\item reset the biases of $A$ and $B$ back to $\eps_0$.
\end{enumerate}
The bias of spin C increases to $2\eps_0(1-2^{-m})$. Thus the biases 
$\left\{2,1,1\right\}$ are asymptotically obtained for the three spins, 
as noted first by Fernandez~\cite{Jose-PhD-Thesis}. This variant, where
$m \gg 1$, is the first exhaustive spin-cooling algorithm. By
generalizing this ``Fernandez algorithm'' 
to more spins, we obtained~\cite{SMW07,EFMW06,EFMW07}
a bias configuration that asymptotically approaches the Fibonacci series; when
all spins are cooled, the following biases are reached:
$\left\{\ldots ,34,21,13,8,5,3,2,1,1\right\}.$
Algorithms based on the 3B-Comp were recently reviewed~\cite{Kaye07}.

%
Following the Fibonacci algorithm, a related algorithm was described, which
involves compression on four adjacent spins~\cite{EFMW06,EFMW07}; The 
tribonacci algorithm reaches biases in accord with the tribonacci series
(also called 3-step Fibonacci series)
$\left\{\ldots ,81,44,24,13,7,4,2,1,1\right\}$, where each term is the sum of 
the three previous terms.
Similarly, one can obtain general $k$-step Fibonacci series, where each term is
the sum of the previous $k$ elements, respectively. Summing over all previous terms produces 
the geometric sequence with powers of two:
$\left\{\ldots ,128,64,32,16,8,4,2,1,1\right\}.$
The AC that reaches this series was termed~\cite{EFMW06,EFMW07}, 
``all-bonacci AC''.
We conclude that, with the same number of spins, exhaustive AC reaches
 a greater degree of cooling than PAC algorithms. For example, the cooling 
factor achieved by Fibonacci with 9 spins is $\FF_{9} = 34$. 
The corresponding enhancement for PAC2 is $(3/2)^4\approx 5$,
and for PAC3 it is about $9\frac{1}{2}$.

%
The \emph{Partner Pairing Algorithm (PPA)} was shown to achieve a 
superior bias than all previous cooling  algorithms, and was proven to be
optimal~\cite{SMW05,SMW07}. The all-bonacci algorithm appears to produce
 the same cooling as the PPA: $\{\ldots,16,8,4,2,1,1\}$ (this was verified
 numerically for small biases~\cite{EFMW06,EFMW07}). In all-bonacci, with $n$
 spins, the 
coldest spin approaches a bias of $\eps_f = 2^{n-2}\eps_0$ (as long as 
$\eps_f \ll 1$). For the PPA, a calculation of the cooling that yields 
$2^{n-2}\eps_0$ was not performed, yet it was proven that 
$2^{n-2}\eps_0 \le \eps_f \le 2^{n-1}\eps_0$. See~\cite{SMW07,SMW05} for a 
proof and see~\cite{EFMW06,EFMW07} for calculations in the case of 
$\eps_f \ll 1$. 

%
The focus of this paper is on simple algorithms that are semi-optimal and also
practicable, owing
to their exclusive reliance on simple logic gates. Such algorithms could be
implemented in the lab if proper physical conditions (e.g., \Tone\ ratios) 
are achieved, and thus they could have practical implications in the near
future. 
In contrast, the optimal algorithms mentioned above, the PPA and the
all-bonacci, do not belong to this category. The all-bonacci requires an
unreasonable number of reset steps~\footnote{The original all-bonacci was
recently found to be much worse than Fibonacci, hence a much improved variant
was designed~\cite{BEMW11}.}, and the permutations required in the sorting step
of the PPA were not yet translated into simple 2-spin and 3-spin gates~\footnote
{An unreasonable number of such simple gates seems to be required; Moreover, the
 translation itself might be extremely resource-intensive~\cite{EFMW06, EFMW07}.
}.

%
The next section describes $\vec{\mathrm{m}}$PAC, a new semi-optimal algorithm
 based on PAC, and its special variant, mPAC, which cool $n=2J+1$
 spins such that the coldest spin is asymptotically cooled by a factor of
 $(2-2^{-m})^J$. The optimal algorithms, the PPA and all-bonacci, need about
 half the spins (more precisely, $J+2$ spins) in order to cool
(asymptotically) the coldest spin to $2^{J}$; we thus see that mPAC is
semi-optimal in the sense that it needs twice as much spins to reach nearly the
same optimal bias. 
Section~\ref{sec:Fib-variants} discusses simple variants of the 
Fibonacci algorithm: $\delta$-Fibonacci~\cite{SMW07, Kaye07}, and mFib, which
 fixes the number
 of cycles at each recursive level. Section~\ref{sec:comparison} compares the
 new algorithms, mPAC and mFib, to
previous cooling algorithms, including the PPA, Fibonacci, all-bonacci, and PAC
algorithms. We show that practicable
versions of mPAC and mFib (with small $m$) reach significant cooling levels
for 5-11 spin systems; Moreover, in the case of a 5-spin system, semi-optimal
cooling levels (half the optimal cooling) are attained with reasonable run-time.
Section~\ref{sec:pure-qubits} provides some analysis 
of SOPAC in case one would like to purify qubits as much as needed in order to
obtain scalable NMRQC\@. We examine the resources required to obtain a highly
 polarized spin.

\section{Semi-optimal PAC-Based Cooling Algorithms}~\label{sec:mPAC-variants}
%
%
The Fernandez and Fibonacci algorithms described above illustrate the improved
level of cooling attainable (asymptotically) by repeated compressions that
involve the target spin. Practicable cooling algorithms that perform a small
number of such cycles (at each recursive level) offer reasonable cooling 
at a reasonable run-time. In this section we describe mPAC, a new PAC-based
algorithm, which approaches nearly half the optimal cooling level
 with only a small number of cycles.

%
The mPAC algorithm that we now present 
is a generalization of PAC2~\footnote{The recent version of PAC2~\cite{EFMW06}
was employed, which is more efficient than the original algorithm~\cite{FLMR04}.
} (see Appendix~\ref{app:alg-PAC} for details on PAC2).
$j\in \{1 \ldots J\}$ is the purification level, $k=2j+1$ is
the bit index, $\BB(k)$ stands for 3B-Comp on spins $(k,k-1,k-2)$ 
which increases the bias of 
$k$, $PT(A\rightarrow B)$ denotes a polarization transfer from bit $A$ to 
bit $B$ (or for simplicity, just a SWAP between their states), 
and $M_0(1)$ is a reset, setting the bias of the reset spin,~$k=1$, to~$\eps_0$.
The algorithm only uses elementary gates: a single gate
operates either on a single spin or on a pair of spins (e.g., PT between
adjacent spins), or on three spins (3B-Comp and PT between next-to-nearest
neighbors).
$M_j(k)$ takes $k=2j+1$ spins at equilibrium and attributes bit $k$ with a bias of $\eps_j$.

\emph{mPAC:}\\
\begin{eqnarray} \label{eq:mPAC-definition}
M_{j}(k) &=& \left[\BB(k) M_{j-1}(k-2)\; PT(k-2\rightarrow k-1)\right.\\
       & & \left.M_{j-1}(k-2)\; \right]^{m} PT(k-2\rightarrow k)\; M_{j-1}(k-2).
       \nonumber
\end{eqnarray}
With three spins
\begin{eqnarray*} 
M_1(3)  =  \left[{\BB}(3)M_0(1) PT(1\rightarrow 2)M_0(1)\right]^{m}PT(1\rightarrow 3)M_0(1).
\end{eqnarray*}
With five spins 
$$M_2(5)=\left[{\BB}(5)M_1(3) PT(3\rightarrow4)M_1(3)\right]^{m}PT(3\rightarrow 5)M_1(3).$$
The recursive formulas above are written from right to left, such that the first
step of $M_j(k)$ is reset of the reset spin, $M_0(1)$, followed by PT from spin
1 to spin 3, and $m$ repetitions of the 4-step sequence in square brackets that
ends with {\em 3B-Comp}. 

%
Notice that the number of spins required by mPAC to achieve a purification level
$J$ is the same as for PAC2:
$n = 2J+1;$ but $\eps_j$ now depends on the number of cycles, $m$: $\eps_j = (2-2^{-m})^j \eps_0$.
For $m=1$ we get back the algorithm PAC2, where $\eps_j = (3/2)^j \eps_0$, thus
1PAC $\equiv$ PAC2; we often retain here the original name, PAC2. 
Asymptotically, for $m \rightarrow \infty$ we get $\eps_j = 2^j \eps_0$.
For three spins, $C,B,A$ and $m>1, $ this is the Fernandez algorithm described
 above, and for sufficiently large $m$, spin $C$ will acquire a final bias of
 $\eps_1 = 2\eps_0$~\cite{EFMW06,EFMW07}. This is calculated via 
$\eps_1 = (2\eps_0+\eps_1)/2 \Rightarrow \eps_1 = 2\eps_0$. 
Similarly, for five spins and sufficiently large $m$, the final bias of the
$5\th$ spin is $\eps_2 = (2\eps_1+\eps_2)/2 \Rightarrow \eps_2 = 2\eps_1 = 4\eps_0$. 

\begin{figure}
\includegraphics[height=0.8\columnwidth,angle=0]{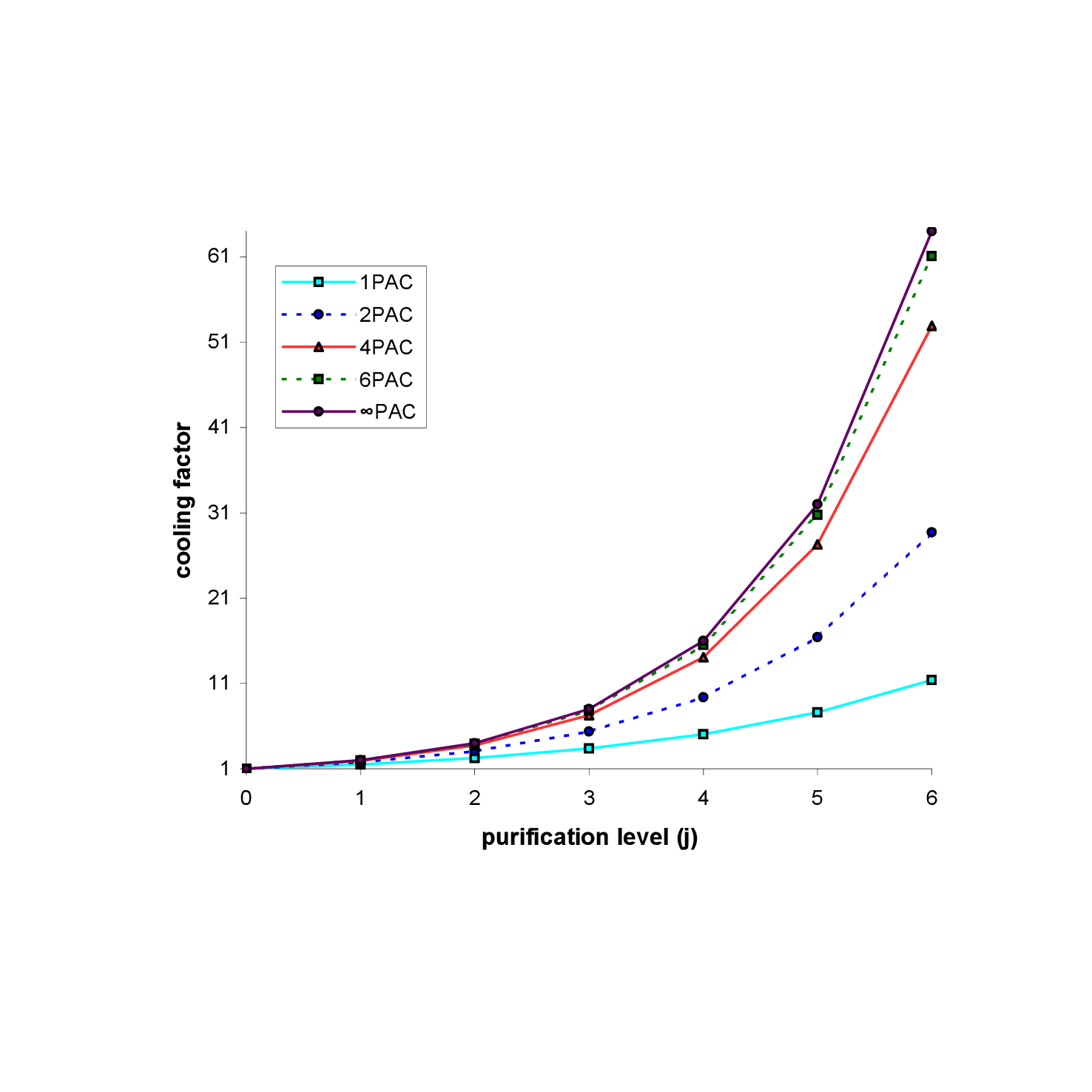}
\caption{Cooling factors ($\eps/\eps_0$, where $\eps_0 =10^{-5}$) for small
spin-systems after mPAC with various $m$ (see Eq.~\ref{eq:mPAC-definition}).}
\label{fig:mPAC-analysis-for-j<7}
\end{figure}

%
The choice of $m$ has a strong influence on the run-time,
but fortunately, the polarization enhancement also increases rapidly with $m$.
For small spin-systems (up to about 10 spins), very small values of $m$ (2--6) 
are sufficient. Figure~\ref{fig:mPAC-analysis-for-j<7} compares the cooling 
factors obtained by such mPAC variants up to 13 spins. Notably, 6PAC cools to a
similar extent as $\infty$-PAC\@. It is also evident that 2PAC cools
significantly better than the single-cycle variant (1PAC $\equiv$ PAC2).

%
%
The run-time of mPAC (neglecting $PT$ and {\em 3B-Comp} steps) is
\begin{eqnarray}\label{eq:mPAC-time-complexity}
T(j)&=& T(j-1) + 2T(j-1)m \\ 
    &=& (1+2m)T(j-1)=(1+2m)^{j},\nonumber
\end{eqnarray} 
in reset-time units~\footnote{
Note that one could use two reset spins instead of one, so that 
$T(j)=T(j-1) + 2T(j-1)m = (1+2m)T(j-1)=(1+2m)^{j-1}T(1)$ with $T(1)=m+1$.
The run-time is therefore reduced by a factor of $\frac{1+2m}{1+m}=
2-\frac{1}{1+m}$.}.
If $m=1$ (mPAC is PAC2) the run-time~\cite{EFMW06} is $T(j)=3^{j}$.
In general, for any integer~$m \ge 1$ 
$$\eps_j=\eps_{j-1}(2-2^{-m})=\eps_0(2-2^{-m})^j.$$ 
For instance, for 21 spins, $j=10$, $m=7$, we get $\eps_j/\eps_0 =985$ 
(as long as the final bias is still much smaller than 1), and for the same $j$,
but $m=4$, $\eps_j/\eps_0 =745 $. For 11 spins, $j=5$, and $m=4$ yields 
$\eps_j/\eps_0=27.3.$

%
In general, for $2^j\eps_0\ll1$ and sufficiently large $m$, spin $k=2j+1$ is 
attributed a bias of $\eps_j=2\eps_{j-1}=2^j\eps_0$.
We denote the asymptotic case ($m \rightarrow \infty$) by ${\infty}$PAC\@.

%
We now generalize mPAC by replacing the constant
$m$ by a vector $\vec{m}$. This added flexibility, in which a different value of
$m$ is associated with each cooling level, was found to be beneficial in the
analysis of the Fibonacci algorithm~\cite{SMW07}, as we explain in
section~\ref{sec:Fib-variants}.
We call this new algorithm $\vec{\mathrm{m}}$PAC; it is defined as follows:\\
\emph{$\vec{m}$PAC:}\\
\begin{eqnarray}\label{eq:vmPAC-recursive}
M_{j}(k)&=&\left[\BB(k)M_{j-1}(k-2)PT(k-2\rightarrow k-1)\right. \\
  && \left. M_{j-1}(k-2) \right]^{m_j} PT(k-2\rightarrow k)\; M_{j-1}(k-2),\nonumber
\end{eqnarray}
%
where (as before) $j$ is the purification level. $M_0(k)$ denotes resetting the
bias of spin $k$ to $\eps_0$.\\
With three spins
\begin{eqnarray*} 
M_1(3)=\left[{\BB}(3)M_0(1) PT(1\rightarrow 2)M_0(1)\right]^{m_1} PT(1\rightarrow 3)M_0(1).
\end{eqnarray*}
With five spins
\begin{eqnarray*} 
M_2(5)=\left[{\BB}(5)M_1(3) PT(3\rightarrow 4)M_1(3)\right]^{m_2}PT(3\rightarrow 5)M_1(3).
\end{eqnarray*}
In general, $M_j(k)$ cools spin $k$ to a bias of $\eps_j=2\eps_{j-1}(1-2^{-m_j}
)$, as $M_j$ is equivalent to performing the Fernandez algorithm on bits $k,k-1,
k-2$, where bits $k-1$ and $k-2$ have equal initial biases $\eps_{j-1}$.
Therefore (see details in Appendix~\ref{app:vmPAC}) 
$$\eps_j=\eps_0\prod_j(2-2^{-m_j}).$$
It is important to mention that $m_j$ could be chosen according to various
criteria; for instance, it could depend on the total number of spins $n$ 
(or on $J$), in addition to its dependence on $j$.
The algorithm requires $\prod_j(1+2m_j)$ reset steps:
\begin{eqnarray}\label{eq:vmPAC-time-complexity}
T(j) & = & T(j-1)+2T(j-1)m_j \\ 
     & = & (1+2m_j)T(j-1)\nonumber 
      =  \prod_{i=1}^j (1+2m_j).
\end{eqnarray} 
Hence, the order in which the $m_j$ appear is irrelevant.

\section{Practicable variants of the Fibonacci algorithm}~\label{sec:Fib-variants}
%
%
The Fibonacci algorithm is exhaustive, in the sense that a very large number of
3B-Comp steps are performed at each recursive level. Practicable variants may be
obtained by limiting the number of compressions, as we describe here. An
algorithm that reaches a distance of
$\delta_{n}^{k}$ from the Fibonacci series was defined as follows:\\ 
\emph{$\delta$-Fibonacci\cite{SMW07}: Run $\FF(n,n)$}
\begin{equation}\label{eq:Fibonacci-definition}
\FF(n , k)=\left[\FF(n,k-1)\; \BB(k)\right]^{m_{n,k}}\FF(n,k-1),
\end{equation}
where $\FF(n,2)$ is a RESET step on bits~1 and~2 (described by $M_0(1) 
PT(1\rightarrow 2) M_0(1)$ in section~\ref{sec:mPAC-variants}), $m_{n,k}$ are
 chosen such that the bias of bit $k$ is at least
 $\FF_{k}(1-\delta_{n}^{k})\eps_0$, and $\FF_k$ is the $k\th$ Fibonacci number. 
Here we choose $\delta_{n}^{k}=2^{k-n-1}$, such that
$\delta_{n}^{n}=\frac{1}{2}$. This condition sets $m_{n,k}=n-k+2$ (see
appendix~\ref{app:fib-run}).

For three spins Fibonacci is 
\begin{eqnarray*}
\FF(3,3) & = & [\FF(3,2)\; \BB(3)]^{m_{3,3}}\FF(3,2)\\\
\FF(3,2) & = & M_0(1)\; PT(1\rightarrow 2)\; M_0(1).
\end{eqnarray*}
This attributes spin number three with a bias of at least 
$
 \FF_3 \left(1-2^{-1}\right)\eps_0 = \eps_0.
$
For four spins 
\begin{eqnarray*}
\FF(4,4) & = & [\FF(4,3)\; \BB(4)]^{m_{4,4}}\FF(4,3)\\\
\FF(4,3) & = & [\FF(4,2)\; \BB(3)]^{m_{4,3}}\FF(4,2)\\\
\FF(4,2) & = & M_0(1)\; PT(1\rightarrow 2)\; M_0(1),
\end{eqnarray*}
which attributes spins three and four with biases of at least 
$\FF_3 \left(1-2^{-2}\right)\eps_0 = \frac{3}{2}\eps_0,$ and
$\FF_4 \left(1-2^{-1}\right)\eps_0 = \frac{3}{2}\eps_0,$ respectively.

%
The term $\delta$-Fib hereinafter always refers to $\delta_n^k = 2^{k-n-1}$, 
such that $m_{n,k}=n-k+2$. For large spin systems, say $n>8$ or so, $\delta$-Fib
 is not practicable, as it requires many cycles in the lower recursion levels.
To circumvent this problem, we fix the number of compression steps, such that
$m_{n,k}=m,$ and denote this variant of Fibonacci by mFib:

\emph{mFib: Run $\FF(n,n)$}\\
\begin{equation} \label{eq:mFib-definition}
\FF(n , k)=\left[\FF(n,k-1)\; \BB(k)\right]^{m}\FF(n,k-1).
\end{equation}
Where $\FF(n,2)$ is a RESET step on bits~1 and~2 as before, and $m$ is any
integer $m \ge 1$.
With three spins
\begin{eqnarray*} 
\FF(3,3)=\left[\FF(3,2)\; \BB(3)\right]^{m}\FF(3,2),
\end{eqnarray*}
and spin three is attributed a bias of
$$\eps_3=(1-2^{-m})(\eps_{1}+\eps_2)=2(1-2^{-m}).$$
With four spins 
\begin{eqnarray*} 
\FF(4,4) & = & \left[\FF(4,3)\; \BB(4)\right]^{m}\FF(4,3)\\\
\FF(4,3) & = & \left[\FF(4,2)\; \BB(3)\right]^{m}\FF(4,2),
\end{eqnarray*}
$\eps_3$ is the same as for three spins, $\eps_{3}=2(1-2^{-m}),$
and spin four is attributed a bias of:
\begin{eqnarray*} 
\eps_{4} & = & (1-2^{-m})(\eps_{2}+\eps_{3})\\\
         & = & (1+2(1-2^{-m}))(1-2^{-m}).
\end{eqnarray*}
In general, for $n$ spins, the bias of the MSB is given by the recursive
formula:
$$\eps_n=(1-2^{-m})(\eps_{n-1}+\eps_{n-2}),$$ where $\eps_1 = \eps_2 = 1.$
Specifically, we focus on cases where $m$ is a small constant ($3-5$). For
$m=2$, it can be shown that mFib is outperformed by PAC3~\cite{BEMW11}.

%
The run-time of mFib is
\begin{eqnarray}\label{eq:mFib-time-complexity}
T(n,m)&=& (m+1)T(n-1) \\
      &=&(m+1)^{n-2}T(2)=2(m+1)^{n-2}.\nonumber
\end{eqnarray} 

\section{Comparison between cooling algorithms}\label{sec:comparison}
%
%
It is interesting to compare SOPAC to other algorithms. We first consider  
the cooling levels attained by each algorithm. ${\infty}$PAC attributes to spin
$n=2J+1$ a bias of $2^{J}\eps_0$.
 In comparison, with $n$ spins, the PPA and all-bonacci reach $2^{n-2}\eps_0$
\footnote{For the PPA, the exact asymptotic bias is not known, however it is
 tightly bounded between $2^{n-2}\eps_0$ and $2^{n-1}\eps_0$; Numerical
 analysis suggests that the lower bound of $2^{n-2}\eps_0$ is the asymptotic
 bias~\cite{EFMW06}.}, and Fibonacci asymptotically approaches $\FF_n\eps_0,$
 where $\FF_n$ is the $n\th$ element of the Fibonacci series.
While the PPA and the Fibonacci algorithm cool the entire spin-system, 
mPAC is defined so that it only polarizes the \emph{most significant bit (MSB)}.
For a fair comparison of run-time we need to cool the entire string in mPAC as
well. To accomplish this, successive applications of mPAC cool the less
significant bits. Namely, the process 
$$M_0(1) PT(1\rightarrow 2) M_0(1) M_1(3) PT(3\rightarrow 4) M_1(3),
\ldots M_J(n)$$  
yields the asymptotic biases
$\{2^{J},2^{J-1},2^{J-1}$ $\ldots$ $,8,8,4,4,2,2,1,1\}.$ 
Consider the application of cooling algorithms to cool all spins (with initial
biases of 0), until the biases are sufficiently close to the
asymptotic biases; the resulting biases of the first seven spins (as long as
 $\eps_f$ of the coldest spin is still much smaller than 1) are given here for
 $\infty$PAC and other exhaustive algorithms, as well as for practicable and
 SOPAC algorithms:
\begin{itemize}
\item $\infty$PAC $\{2^{J},\ldots,8,4,4,2,2,1,1\}$
\item PPA and all-bonacci $\{2^{n-2},\ldots,32,16,8,4,2,1,1\}$
\item Fibonacci $\{\FF_{n},\ldots,13,8,5,3,2,1,1\}$
\item PAC2 $\equiv$ 1PAC $\left\{ (3/2)^J \ldots, 3\frac{3}{8}, 2\frac{1}{4},
2\frac{1}{4}, 1\frac{1}{2}, 1\frac{1}{2}, 1,1\right\}$
\item PAC3 $\left\{ \ldots, 5\frac{1}{8},3\frac{3}{4},2\frac{3}{4},2,1\frac{1}{2},1,1\right\}$
\item 2PAC $\left\{ \ldots, 5.36,3.06,3.06,1.75,1.75,1,1\right\}$
\item 4PAC $\left\{ \ldots, 7.27,3.75,3.75,1.94,1.94,1,1\right\}$
\item 6PAC $\left\{ \ldots, 7.81,3.94,3.94,1.98,1.98,1,1\right\}$
\item 3Fib $\left\{ \ldots, 7.81,5.29,3.64,2.41,1.75,1,1\right\}$
\item 4Fib $\left\{ \ldots, 10.2,6.54,4.28,2.70,1.88,1,1\right\}$
\end{itemize}
The bias configurations are given in units of the initial bias, $\eps_0$.
$\delta$-Fib is not included, as for each total number of spins it
produces different bias series. For $n=7, \delta$-Fib yields the biases $\{8.27,
6.49,4.54,2.88,1.97,1,1\}$.

%
We consider a small spin-system comprising five spins.
Table~\ref{tab:bias-5-spins} compares the biases (for the MSB) obtained by
previous algorithms (top), mPAC (middle section), and mFib (bottom), as well
as the number of resets required to create the entire bias series~\footnote{
Ignoring the other steps is equivalent to assuming that any $n$-bit gate is
performed in a single computing step, and that the total number of such
computing steps is negligible with respect to the duration of the reset steps.
In reality this is not the case, and practicable algorithms such as PAC and
SOPAC are important.}.
Note that the PPA cools better when more resets are allowed, approaching the
limit of $8\eps_0$ for AC with 5 spins. With only 28 resets, the PPA attains
a semi-optimal cooling level of $4\eps_0$; For a similar cooling level, 4PAC and
6PAC require 101 and 197 resets, respectively. After 99
resets, the PPA obtains a near-optimal cooling level of $7\eps_0$.
Table~\ref{tab:bias-7-spins} presents a similar comparison for 7 spins.
Table~\ref{tab:bias-9-spins} and Table~\ref{tab:bias-11-spins} present similar
comparisons for 9 and 11 spins, respectively; a spin-system of comparable size
was recently used for benchmarking quantum control methods~\cite{NMR+06}.
Appendix~\ref{app:cool-factors} compares the number of spins and run-time
required by each algorithm to achieve several small cooling factors.

%
The run-time analysis of the PAC algorithms is conveniently expressed as a 
function of the purification level, $J$ (see Appendix~\ref{app:alg-PAC}).
The entire spin-system is cooled by $2J-1$ successive applications of each
algorithm, as shown above for mPAC.
In contrast, the Fibonacci algorithm and the PPA were
 designed to generate the entire series of biases. The run-time of
 $\delta$-Fib is given by $n!$
(see calculation in appendix~\ref{app:fib-run}), and the run-time of the PPA
 was obtained by a computer simulation that iterates between the two steps
 of the algorithm.

\begin{table}[here]
{\centering
 \begin{tabular}{l|c|c}
&
bias ($\eps_0$) &
\parbox{13ex}{Run-time \\ (reset steps)}
\\\hline\hline
$\delta$-Fib & 3.29  & 120
\\
PAC2 & 2.25  & 17
\\
PAC3      & 2.75  & 29 
\\
PPA      & 4.03  & 28
\\
PPA      & 7.00 & 99 
\\\hline
2PAC      & 3.06  & 37
\\
4PAC      & 3.75  & 101
\\
6PAC      & 3.94  & 197
\\\hline
3Fib      & 3.64  & 128 
\\
4Fib      & 4.28  & 250 

\end{tabular}\par}\ \\
\caption{%
 Cooling a 5-spin system by various algorithms. The optimal cooling level is $8\eps_0$.
}
\label{tab:bias-5-spins}
\end{table}

\begin{table}[here]
{\centering
 \begin{tabular}{l|c|c}
&
bias ($\eps_0$) &
\parbox{13ex}{Run-time \\ (reset steps)}
\\\hline\hline
$\delta$-Fib & 8.27  & 5040
\\
PAC2 & 3.38  & 53
\\
PAC3      & 5.13  & 169 
\\
PPA      & 8.02 & 104
\\
PPA      & 16.0  & 428
\\\hline
2PAC      & 5.36  & 187
\\
4PAC      & 7.27  & 911
\\
6PAC      & 7.81  & 2563 
\\\hline
3Fib      & 7.81  & 2048 
\\
4Fib      & 10.15  & 6250 

\end{tabular}\par}\ \\
\caption{%
 Cooling a 7-spin system by various algorithms. The optimal cooling level is $32\eps_0$.
}
\label{tab:bias-7-spins}
\end{table}

\begin{table}[here]
{\centering
 \begin{tabular}{l|c|c}
&
bias ($\eps_0$) &
\parbox{13ex}{Run-time \\ (reset steps)}
\\\hline\hline
$\delta$-Fib & 21.5  & 362880
\\
PAC2 & 5.06  & 161
\\
PAC3      & 9.56  & 985
\\
PPA      & 32.0 & 1639
\\
PPA      & 64.0 & 6836 
\\\hline
2PAC      & 9.38 & 937 
\\
4PAC      & 14.1 & 8201 
\\
6PAC      & 15.5 & 33321 
\\\hline
3Fib & 16.9 & 32768 
\\
4Fib & 24.2 & 156250
 
\end{tabular}\par}\ \\
\caption{%
 Cooling a 9-spin system by various algorithms. The optimal cooling level is 
 $128\eps_0$. 
}
\label{tab:bias-9-spins}
\end{table}

\begin{table}[here]
{\centering
 \begin{tabular}{l|c|c}
&
bias ($\eps_0$) &
\parbox{13ex}{Run-time \\ (reset steps)}
\\\hline\hline
$\delta$-Fib & 56.0  & 39916800
\\
PAC2 & 7.59  & 485
\\
PAC3      & 17.8  & 5741
\\
PPA      & 64.0 & 6456
\\
PPA      & 256 & 109323 
\\\hline
2PAC      & 16.4 & 4687 
\\
4PAC      & 27.3 & 73811 
\\
6PAC      & 30.8 & 433175 
\\\hline
3Fib & 36.4 & 524288 
\\
4Fib & 57.7 & 3906250 
 
\end{tabular}\par}\ \\
\caption{%
 Cooling an 11-spin system by various algorithms. The optimal cooling level is 
 $512\eps_0.$ 
}
\label{tab:bias-11-spins}
\end{table}

\section{Purification of qubits}\label{sec:pure-qubits}
%
%
The practicable nature of the algorithmic cooling described in the previous
sections might hold significant potential for deriving scalable NMR quantum
computation. The pseudopure states used in NMRQC were found to scale well for
sufficiently pure spins, for which $\eps = 1 - 2\delta$, where the error
probability, $\delta, $ satisfies $\delta \ll 1,$ see~\cite{BMRVV02}. For small spin
systems, $\delta \sim 0.2$ is already useful~\footnote{Several spins must be
60\% polarized in order to enable scalable quantum computing based on pseudopure states.},
while larger systems require
($\delta$ of the order of $1/n$)~\cite{BMRVV02}. Evaluation of cooling
algorithms for qubit purification requires exact calculations that go beyond the
approximation of $\eps \ll 1$, which was used so far. Here we describe the 
exact result of 3B-Comp, and use it in the general analysis of mPAC\@.

%
The precise result of 3B-Comp is given by~\cite{EFMW06,EFMW07}:
$$\eps'_3=\frac{\eps_3+\eps_2+\eps_1-\eps_3\eps_2\eps_1}{2}.$$ 
Performing 3B-Comp iteratively while replenishing $\eps_1$ and $\eps_2$ 
after each compression will give, in the $m^{th}$ iteration:
\begin{eqnarray*}
\eps_3^{(m)} & = & \frac{\eps^{(m-1)}_3+\eps_2+\eps_1-\eps_3^{(m-1)}\eps_2
                   \eps_1}{2}\\
             & = &\frac{\eps_2+\eps_1+\eps_3^{(m-1)}(1-\eps_2\eps_1)}{2}.
\end{eqnarray*}
%

%
In the case of mPAC, $\eps_1=\eps_2$ and therefore
\begin{equation}\label{eq:eps_3}
\eps_3^{(m)}=\frac{\eps_3^{(m-1)}(1-\eps_1^2)}{2}+\eps_1.
\end{equation}
Defining 
\begin{equation}\label{eq:defining-A}
A(\eps)\triangleq\frac{1-\eps^2}{2},
\end{equation} 
we solve Eq.~\ref{eq:eps_3}.
\begin{eqnarray*}
\eps_3^{(m)} 
   & = & A(\eps_1)\eps_3^{(m-1)}+\eps_1 \\
   & = & A(\eps_1)^2\eps_3^{(m-2)}+(A(\eps_1)+1)\eps_1 \\
   & = & A(\eps_1)^{m}\eps_3^{(0)}+\eps_1\sum_{i=0}^{m-1}A(\eps_1)^i.
\end{eqnarray*}
In our scheme, $\eps_3^{(0)}=\eps_1$ and so 
\begin{equation}
\eps_3^{(m)} 
   = \eps_1\sum_{i=0}^{m}A(\eps_1)^i 
   = \eps_1\frac{1-A(\eps_1)^{1+m}}{1-A(\eps_1)},
\end{equation}
which can be generalized to the following recursive formula:
\begin{equation}\label{eq:eps_k(m)}
\eps_k^{(m)}  
   = \eps_{k-2}\sum_{i=0}^{m}A(\eps_{k-2})^i 
   = \eps_{k-2}^{(m)}\frac{1-A\left(\eps_{k-2}^{(m)}\right)^{1+m}}
                    {1-A\left(\eps_{k-2}^{(m)}\right)}.
\end{equation}
%

\begin{figure}
\includegraphics[height=0.8\columnwidth,angle=0]{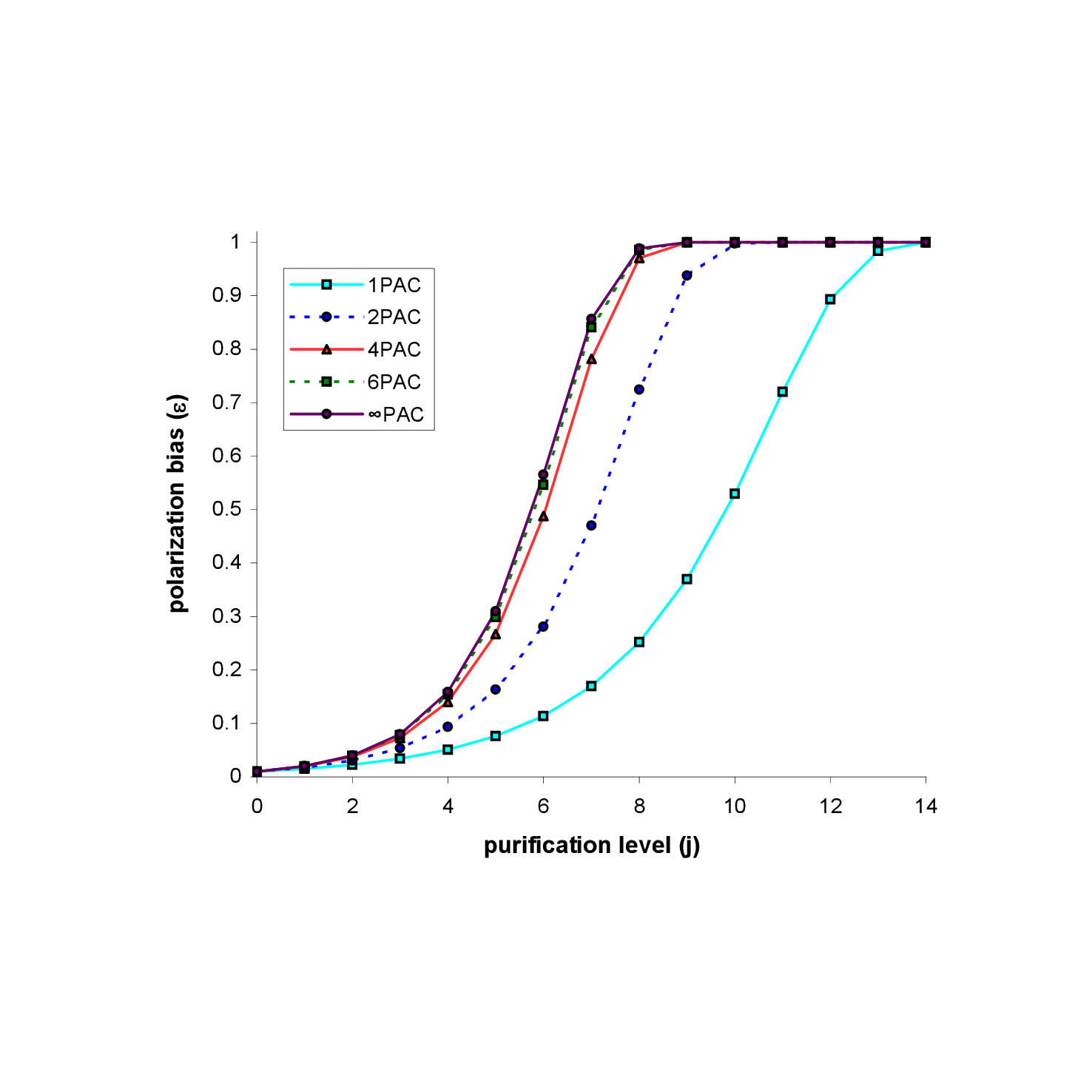}
\caption{Polarization biases ($\eps_{2j+1}$) for mPAC with various $m$ as a
function of purification level $j$ (defined in Eq.~\ref{eq:mPAC-definition}).
The initial bias is $\eps_0=0.01$. The curve for $\infty$PAC (i.e.,
 $m\rightarrow\infty$) was calculated from Eq.~\ref{eq:inftyPAC-recursive-eq}.}
\label{fig:mPAC-analysis-for-0.01} 
\end{figure}

\begin{figure}
\includegraphics[height=0.8\columnwidth,angle=0]{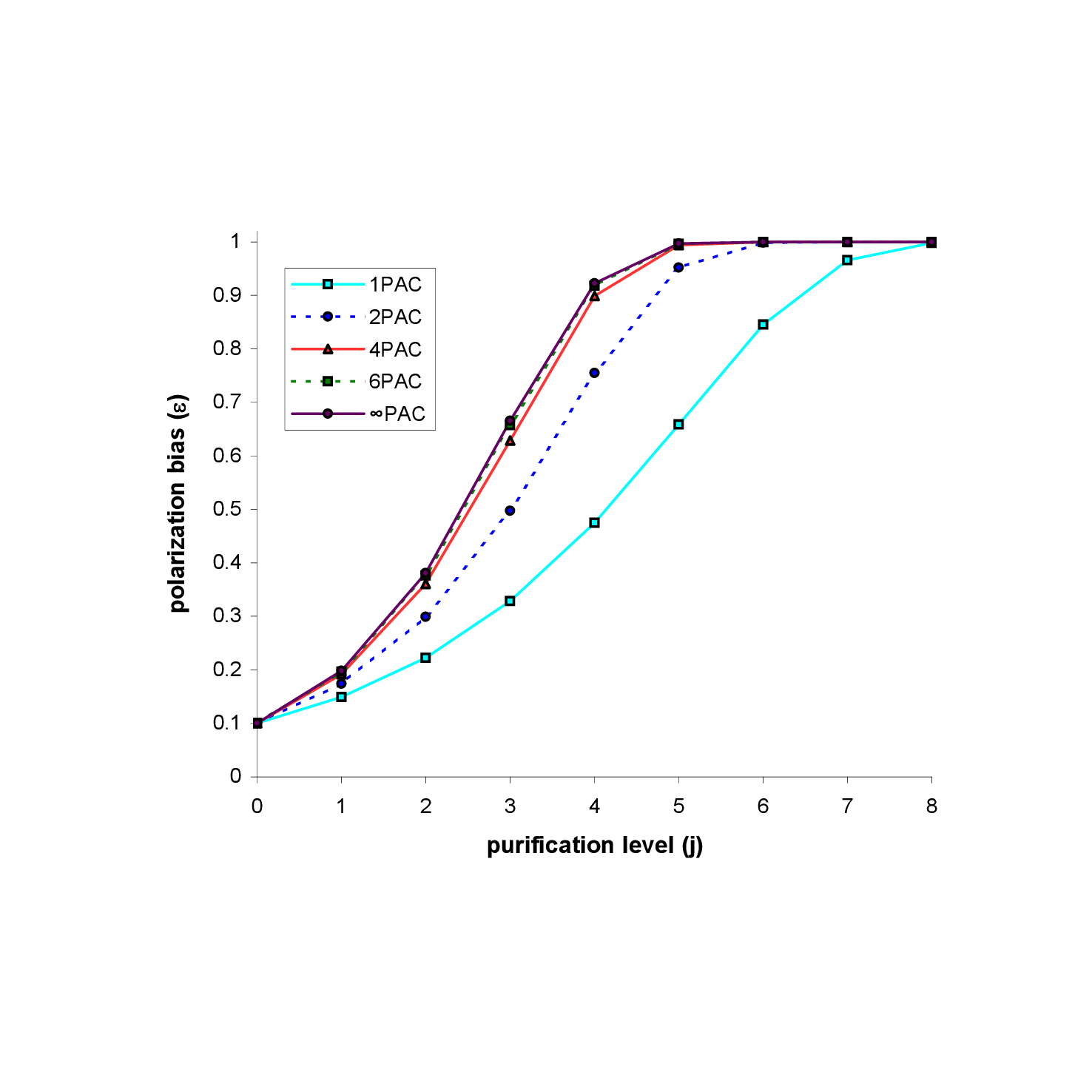}
\caption{Polarization biases ($\eps_{2j+1}$) for practicable mPAC variants as a
 function of purification level $j$. The initial bias is $\eps_0=0.1$.}
\label{fig:mPAC-analysis-for-0.1} 
\end{figure}

%
Cooling curves for several mPAC variants are presented in
Figure~\ref{fig:mPAC-analysis-for-0.01}, starting from a typical electron 
polarization of 1\%. The curve for 6PAC is similar to $\infty$-PAC (see
Figure~\ref{fig:mPAC-zoom-for-0.01} in Appendix~\ref{app:mPAC-zoom} for a
 close-up view); 6PAC
requires about 13 spins to attain a semi-optimal bias, and with about 17
spins it reaches polarization near unity. Starting from hyperpolarization
($\eps_0 = 10\%$), 6PAC exceeds 50\% polarization with only 7 spins (see
Figure~\ref{fig:mPAC-analysis-for-0.1} and Figure~\ref{fig:mPAC-zoom-for-0.1});
 In this case, 11 spins are sufficient to purify one spin to a high degree.

%

%
Table~\ref{tab:compare-0.6} gives the resources required by SOPAC in order to 
reach the threshold of scalable NMRQC, $\eps = 0.6$~\cite{BMRVV02}; The left
pane assumes an
initial polarization of 1\%, a typical electron polarization, while the right
pane assumes $\eps_0 = 10\%$. Results for mFib were obtained numerically.
Note that for $m=6$, mPAC requires the same amount of spins and 
more reset steps than 4PAC, and produces slightly higher biases.

\begin{table}[ht]
\begin{center}
\begin{tabular}{|c||c|c|c||c|c|c|}
\cline{2-7}
 & \multicolumn{6}{|c|}{$\eps\ge0.6$}\\
\cline{2-7}
 & \multicolumn{3}{c|}{$\eps_0=0.01$} & \multicolumn{3}{c|}{$\eps_0=0.1$}\tabularnewline
\hline
\hline 
alg. & \#spins & run-time & bias & \#spins & run-time & bias\tabularnewline
\hline 
1PAC & 23 & 354293 & 0.72 & 11 & 485 & 0.66\tabularnewline
\hline 
2PAC & 17 & 585937 & 0.72 & 9 & 937 & 0.76\tabularnewline
\hline 
4PAC & 15 & 5978711 & 0.78 & 7 & 911 & 0.63\tabularnewline
\hline\hline
3Fib & 13 & 97656250 & 0.72 & 7 & 2048 & 0.67\tabularnewline
\hline 
4Fib & 12 & 19531250 & 0.72 & 7 & 6250 & 0.78\tabularnewline
\hline 
\end{tabular}
\caption{\label{tab:compare-0.6}
Number of spins and run-time required by SOPAC to purify the MSB to a
polarization of at least 60\%, starting from an initial polarization of 1\%
(left) or 10\% (right). For the purification from 1\%, 3PAC requires 15 spins, 
like 4PAC, and less resets (1098057), but reaches a lower purification (68\%).
}
\end{center}
\end{table}

\subsection*{Fault-tolerant quantum computing}
%
%
For the purpose of fault tolerant quantum computing, it is essential to obtain
highly pure spins, where the error probability, $\delta = \frac{1-\eps}{2}, $ is
well below 1\%. Table~\ref{tab:compare-0.9999} shows the number of resets and
spins required by mPAC variants to achieve $\eps\geq 0.9999$ for the MSB, 
starting from $\eps_0=0.1$, which is a typical electron spin polarization at low
 temperature. For $m=6, $ mPAC still requires 13 spins and more reset steps
 than 4PAC (the bias achieved by 6PAC is $0.99999$).

%
It is notable that 2PAC and 4PAC achieve such high purity with only six more
 spins (three additional purification levels) with respect to the minimal
 purification ($\sim 60\%-80\%$ purity) obtained in Table~\ref{tab:compare-0.6}.
This rapid purification stems from the behavior of mPAC at low error
 probabilities, which may be obtained by substituting $\eps$ by $1-2\delta$ in
 Eq.~\ref{eq:eps_k(m)}, and considering the first three terms of the summation
 ($i=0,1,2$):
$$1-2\delta_{j+1} = (1-2\delta_j)\left[1+2\delta_j(1-\delta_j)+
4\delta_j^2(1-\delta_j)^2\right],$$
where $\delta_j$ is the error probability at purification level $j,$ and 
$A(\eps)=A(1-2\delta)=2\delta(1-\delta)$ (see Eq.~\ref{eq:defining-A}). 
At low error probabilities, $\delta_{j+1} \xrightarrow{\delta_j\ll1}\delta_j^2$.
As a result of this rapid convergence, the number of spins required by mPAC to
 reach 99.9\% purity is sufficient to produce a markedly higher purity of
 99.99\% (or even 99.999\% for some variants). 

%
Table~\ref{tab:compare-0.9999} includes two mFib SOPAC variants, which require
fewer spins than the practicable mPAC variants; Both 3Fib and 4Fib require 11
 spins, while mPAC requires at least 13 spins. The run-times and cooling levels
 of 3Fib and 4PAC are similar, while 4Fib has a much higher run-time and
 reaches extremely high purification (99.9999\%).

%

%
Obviously, only the case of purifying from $\eps=0.1$ to above 60\% polarization
can be considered as a potential near future application. The other cases
presented here require an unreasonable run-time.

\begin{table}[ht]
\begin{center}
\begin{tabular}{|c||c|c|c|}
\hline
\hline 
alg. & \#spins & run-time & final bias \tabularnewline
\hline 
1PAC & 19 & $39365$ & 0.999996\tabularnewline
2PAC & 15 & $117187$ & 0.999999\tabularnewline
4PAC & 13 & $664301$ & 0.999984\tabularnewline
\hline 
3Fib & 11 & $524288$ & 0.999877\tabularnewline
4Fib & 11 & $3906250$ & 0.999999 \tabularnewline
\hline
\end{tabular}
\caption{\label{tab:compare-0.9999}
Number of spins and run-time required by SOPAC to achieve a minimum bias of
$0.9999$, starting from $\eps_0=0.1$. With the same number of spins as 4PAC,
3PAC reaches a slightly lower bias ($0.999938$) using fewer resets (156865).
}
\end{center}
\end{table}

\section{Discussion}
%
%
We have introduced SOPAC, semi-optimal practicable cooling algorithms that
achieve better cooling than the PAC algorithms by performing a small number of
cycles at each recursive level. The $\vec{\mathrm{m}}$PAC and its special
instance mPAC (constant $m$) offer improved cooling for small spin systems at a
 reasonable increase in run-time. Similar results were obtained by means of the
 mFib algorithm, which requires a constant number ($m$) of 3-bit compressions at
 each
 recursive level. We compared the performance of mPAC and mFib (with small
 $m$) with the previous algorithms - the PAC algorithms (PAC2 and PAC3), the
 PPA, and $\delta$-Fib. While the PPA is proven to be optimal,
 it is not expected to be the first choice in practice, since an efficient
 implementation of its SORT step might not be found. The SOPAC algorithms allow
 flexible tuning of the degree of cooling, thereby bridging the large gap
 between the elementary cooling of PAC, and the extensive, yet unattainable, 
 cooling of exhaustive algorithms.

%
AC was mentioned as an enabling technique for various quantum computing
 schemes~\cite{Twamley03,FS03,LGYY+02,PBT10}. The simplicity of implementation and
 semi-optimal nature of SOPAC renders it potentially useful for deriving
 spin-based quantum computers in the more distant future~\cite{BMRVV02,
 Twamley03, FS03,LGYY+02,PBT10}. To this end, the new practicable cooling
 algorithms
 presented here may yield sufficiently pure qubits, starting from reasonably
 small spin systems initially polarized by nearby electrons at room
 temperature ($\eps_0=0.01$) or at low temperature ($\eps_0=0.1$). In the second
 case, both 3Fib and 4PAC require only seven spins to cool the MSB beyond 60\%
 polarization, permitting scalable NMRQC~\cite{BMRVV02}. With about five additional spins,
 these algorithms can ideally produce highly purified spins ($\eps\ge 0.9999$)
 that are suitable for fault tolerant quantum computing. While a similar 12-spin
 system was recently manipulated in liquid-state NMR, more spins would be
 required in practice due to various factors, such as actual thermalization
 times and decoherence in between reset steps due to spin-spin relaxation.

%
AC may contribute to NMR spectral editing, wherein part of the spectrum 
that corresponds to a particular spin of interest is selectively enhanced with
respect to overlapping signals. In such cases, SOPAC can potentially offer
 selective enhancement much beyond simple polarization transfer.

%
Last but not least, in real life, the reset spins do not relax 
infinitely faster than the computation spins (see~\cite{AGN04,RMM07,HRM07,
WHRSM08,BP08,LPS10} for viewing AC as a novel type of heat-engine).
Let ${\cal R}$ denote the ratio between the relaxation time of the 
computation spins and the relaxation time of the reset spins, 
${\cal R}= \Tone(\textrm{comp.})/\Tone(\textrm{reset})$.
Ref~\cite{BEMW11} thoroughly analyzes various algorithms, including 2PAC 
suggested here, for several values of ${\cal R}$, and found (as expected)
that when the ratio is much larger than the run-time (number of reset steps)
required by the algorithm, ${\cal R} \gg T_{run-time},$ the 
results here still apply with only minor corrections. However, when ${\cal R}$
is in the range of about $T_{run-time}$ to $10T_{run-time},$ the deviations
 become significant.
For example, Ref~\cite{BEMW11} considers the application of 2PAC to 7 spins,
 starting
 from $\eps_0 \ll 1.$ When ${\cal R}=10000$ and the duration of each reset step
is $5\Tone(\textrm{reset})$, cooling all the spins attributes the MSB with a
 final bias of $5.11$ (in units of $\eps_0$), slightly below
 the ideal bias ($5.36$, see Table~\ref{tab:bias-7-spins}); lower ratios,
 ${\cal R}= 1000$ and ${\cal R}= 100,$ significantly reduce the final bias, to
 $3.63$ and $1.07$, respectively. When only the MSB is cooled, the same bias,
 $5.36,$ is reached in the ideal case, while the run-time is shortened from 187
to 125 reset steps, hence significantly higher biases are obtained; for
 ${\cal R}=10000\ ({\cal R}=1000, {\cal R}=100)$ the resulting biases are $5.24$
(respectively $4.58, 2.56$). 
The final bias may be further increased by optimizing the duration of the 
reset step (in particular for small ${\cal R}$); For example, when
 ${\cal R}=100$, reducing the reset duration to $1.8\Tone(\textrm{reset})$ 
increases the final bias from $2.56$ to $3.05,$ when cooling only the MSB,
while reducing the reset period further, to $0.91\Tone(\textrm{reset})$, nearly
 doubles the final bias, from $1.07$ to $1.97$, when all spins are cooled
 (see~\cite{BEMW11} for more details). Notably, ${\cal R} \sim 10000$ was
 achieved in solid-state NMR, where spin-diffusion was used for rapid
 repolarization~\cite{BMR+05, RMB+08}.

\section*{Acknowledgements}
%
%
This work was supported in part by the Wolfson Foundation, and by the
 Israeli MOD Research and Technology Unit. The work of T.M. was also supported
in part by FQRNT through INTRIQ, and by NSERC.

%
%

%
%
%
%
%
\clearpage 
\noindent{\LARGE \bf Appendices}
\appendix

%
\section{Formal descriptions of PAC algorithms}\label{app:alg-PAC}
PAC2 on $2J+1$ spins can reach a purification level $J$ on the MSB spin ($k=2J+1$)~\cite{FLMR04}.
The procedure $M_0(1)$ is defined as a reset step on spin $k=1$.
The procedure $M_j(k)$ is defined recursively to cool the $k\th$ spin to a 
purification level $j \in {1 \ldots J}$ (i.e. $\eps_j=(3/2)^j$). 
$\BB(k)$ denotes 3B-Comp on spins $(k,k-1,k-2)$ 
which increases the bias of $k$ from level $j-1$ to level $j$, 
and $PT(A\rightarrow B)$ denotes a polarization transfer from bit $A$ to bit $B$. 

\emph{Practicable algorithmic cooling 2 (PAC2):\\
For $j\in \{1,\ldots ,J\}$
\begin{eqnarray*}
M_{j}(k)&=&\BB(k)
M_{j-1}(k-2)PT(k-2\rightarrow k-1)\\
        & & M_{j-1}(k-2)PT(k-2\rightarrow k)M_{j-1}(k-2),\nonumber
\end{eqnarray*}
}
\\
With $M_0(1)$ being reset on spin 1. For example, to cool spin 3 to the first level of purification we use:
\begin{eqnarray*}
M_{1}(3) & = & \BB(3)M_{0}(1)\\
         &   & PT(1\rightarrow 2) M_{0}(1)PT(1\rightarrow 3) M_{0}(1),
\end{eqnarray*}
and to cool spin~5 to the second purification level we use
\begin{eqnarray*}
M_{2}(5)&=&\BB(5)M_{1}(3)\\
        & & PT(3\rightarrow 4) M_{1}(3)PT(3\rightarrow 5) M_{1}(3).
\end{eqnarray*}
PAC2 requires $J=\left\lceil \log_{1.5}(\eps_{J}/\eps_0)\right\rceil,$ as every
 3B-Comp step boosts the bias ($\eps_j \ll1$) by a factor of 1.5~\cite{FLMR04}. 
The calculated run-time of PAC2 is $3^{J},$ for cooling the MSB~\cite{FLMR04}.

PAC3 is described as follows~\cite{EFMW06}: 
\\
\emph{Practicable algorithmic cooling 3 (PAC3):\\
for $k\in \{2,\ldots ,J\}$
(the reset spin is at index~1)
\begin{eqnarray*}
M_{k}(k+1) & = & \BB(k+1)
 M_{k-2}(k-1) \\ 
           &   & M_{k-1}(k) PT(k\rightarrow k+1) M_{k-1}(k),
\end{eqnarray*}
}
\\
with $M_0(1) = {\rm RESET}$ (of the reset spin), and \newline $M_1(2) =
PT(1\rightarrow 2) M_0(1)$. For three spins PAC3 is expressed as
$$M_{2}(3)=\BB(3) M_{0}(1)\; M_{1}(2)\; PT(2\rightarrow 3) M_{1}(2) ,$$
and with four spins PAC3 is described by 
$$                                                                     
M_{3}(4)=\BB(4)                                                                      
M_{1}(2)\; M_{2}(3)\; PT(3\rightarrow 4) M_{2}(3).
$$
The resulting bias series on 7 spins is 
$$
\left\{5\frac{1}{8},3\frac{3}{4},2\frac{3}{4},2,\frac{3}{2},1,1\right\}.
$$
The run-time of PAC3 is given by the recursive formula~\cite{EFMW06}:
$T_J=2T_{J-1}+T_{J-2},$
where $T_0=T_1=1,$ and $J=n-1.$


\section{Analysis of $\vec{\mathrm{m}}$PAC}\label{app:vmPAC} 
We can generalize the expression for the achieved bias (Eq.~\ref{eq:eps_k(m)}) 
to \begin{equation}\label{eq:recursive}
\eps_k^{(m_j)}  
   = \eps_{k-2}^{(m_{j-1})}\frac{1-A\left(\eps_{k-2}^{(m_{j-1})}\right)^{1+m_j}}
                    {1-A\left(\eps_{k-2}^{(m_{j-1})}\right)}.
\end{equation}
The limit of this algorithm is given by
\begin{equation}\label{eq:inftyPAC-recursive-eq}
\lim_{\forall j, m_j\rightarrow\infty}\eps_k^{(m_j)}  
   = \frac{\eps_{k-2}^{(\infty)}}
                    {1-A\left(\eps_{k-2}^{(\infty)}\right)},
\end{equation}
where $\eps_{1}^{(\infty)}=\eps_{2}^{(\infty)}=\eps_0.$

In the limit of $\eps_{k}\ll1$ we can use the first order approximation in
$\eps_{k}$. The bias assigned to the target bit, $k$, thus depends only on
$\{m_i|i=1,\ldots j\}$ (i.e., with no significance to the order of appearance). 
For example, $\vec{m}$PAC on five spins which uses  
$m_1=3,m_2=5$ will set the coldest bit to the same bias as the sequence 
$m_1=5,m_2=3$. To prove this let us first notice that Eq.~\ref{eq:defining-A} 
becomes $A(\eps)\xrightarrow{\eps\ll1}\frac{1}{2}$.  
Eq.~\ref{eq:recursive} then has a simple analytical solution
\begin{equation}
\eps_k^{(m_j)} = 
    2\eps_{k-2}^{(m_{j-1})} \left(1-2^{-1-m_j}\right)=
    2^j\eps_0\prod_{i=1}^{j}\left(1-2^{-1-m_i}\right).
\end{equation}

%
\section{Derivation of the run-time of $\delta$-Fib}\label{app:fib-run} 
The choice of $m_{n,k}=n-k+2$ in Eq.~\ref{eq:Fibonacci-definition} yields 
biases that approach the Fibonacci series ($\FF_{n}, \FF_{n-1}, \ldots$), up to
a factor of $1-\delta_{n,k} = 1- 2^{k-n-1}$.
We assume by induction that bits $k-1$ and $k-2$ attained the biases 
$\eps_0a_1>\eps_0\FF_{k-1}(1-2^{k-n-2})$ and
$\eps_0a_2>\eps_0\FF_{k-2}(1-2^{k-n-3})$, respectively. We shall now prove that 
bit $k$ will attain a bias~$\eps_0x>\eps_0\FF_{k}(1-2^{k-n-1})$. 

The effect of the 3B-Comp gate applied to the bits $k-2,k-1,k$ with 
the corresponding biases $\eps_0 a_1,\eps_0 a_2$, and $\eps_0 x$, assigns 
bit $k$ the bias $\eps_0 x\rightarrow\eps_0 (a_1+a_2+x)/2$ (with $a_1\eps_0,
a_2\eps_0\ll 1$). 
In each iteration within the $k\th$ recursive level, $x$ converges towards the
unique fixed point of this transformation, $x = a_1+a_2$. The convergence is
 rapid; after $m_{n,k}$ iterations, starting at $x=0$, Fibonacci achieves
 $x=(a_1+a_2)(1-2^{-m_{n,k}}).$
Note that 
\begin{eqnarray*}
a_1+a_2 & > &\FF_{k-1}(1-2^{k-n-2}) + \FF_{k-2}(1-2^{k-n-3})\\
	& = &\FF_k-2^{k-n-3}(2\FF_{k-1}+\FF_{k-2})\\
        & = &\FF_k-2^{k-n-3}\FF_{k+1} > \FF_k(1-2^{k-n-2}),
\end{eqnarray*}
where the last inequality is due to $\FF_{k+1} < 2\FF_{k}$.
Therefore, by choosing $m_{n,k}=n-k+2$, the Fibonacci algorithm will achieve 
\begin{eqnarray*}
x & > & \FF_k(1-2^{k-n-2})(1-2^{-m_{n,k}})=\FF_k(1-2^{k-n-2})^2\\
  & > & \FF_k(1-2^{k-n-1}), 
\end{eqnarray*}
where the last line is due to $(1-a)^2 > 1-2a, $ for any positive $a$.
\qed

According to the definition of Fibonacci in Eq.~\ref{eq:Fibonacci-definition}, 
the run-time $T(n,k)$ of the operation $\FF(n,k)$ is given by 
\begin{eqnarray*}
T(n,k) & = & (m_{n,k}+1)T(n,k-1)\\
         & = & T(n,2)\prod_{j=3}^k (m_{n,k}+1)=2\prod_{j=3}^k (n-j+3),
\end{eqnarray*}
where only the run-time of RESET steps is taken into account. $T(n,2)=2$ because
$\FF(n,2)$ involves two RESET steps.
Therefore, the run-time of the full Fibonacci algorithm, $\FF(n,n), $ is:
$$T(n,n)=2\prod_{j=3}^n (n-j+3)=2\prod_{i=3}^{n}i=n!.$$
Only half the run-time is required if the two reset steps in $\FF(n,2)$ are
 performed in parallel. 

%
\section{Comparison of algorithms for small target cooling factors}
\label{app:cool-factors}
It is interesting to compare the performance of SOPAC with PAC and other 
cooling algorithms for fixed target cooling factors.
Tables~\ref{tab:biases>=3},~\ref{tab:biases>=7},~\ref{tab:biases>=11},
and~\ref{tab:biases>=15} list the total number of spins and reset steps required
by each cooling algorithm to reach biases of at least $3\eps_0, 7\eps_0, 
11\eps_0, $ and $15\eps_0$, respectively.

\begin{table}[here]
{\centering
 \begin{tabular}{l|c|c}
&
\#spins &
\parbox{13ex}{Run-time \\ (reset steps)}
\\\hline\hline
$\delta$-Fib & 5  & 120 
\\
PAC2 & 7  & 53 
\\
PAC3      & 6  & 70 
\\
PPA      & 4  & 16 
\\\hline
2PAC      & 5 & 37
\\\hline
3Fib & 5 & 128 
\\
\end{tabular}\par}\ \\
\caption{%
A comparison of various algorithms for a target bias of at least $3\eps_0$.
The only SOPAC variants shown are 2PAC and 3Fib, as other practicable variants
 require the same amount of spins and more reset steps.
}
\label{tab:biases>=3}
\end{table}

\begin{table}[here]
{\centering
 \begin{tabular}{l|c|c}
&
\#spins &
\parbox{13ex}{Run-time \\ (reset steps)}
\\\hline\hline
$\delta$-Fib & 7  & 5040
\\
PAC2 & 11 & 485
\\
PAC3      & 8  & 408
\\
PPA      & 5  & 97 
\\\hline
2PAC      & 9 & 937 
\\
4PAC      & 7 & 911 
\\
6PAC      & 7 & 2563
\\\hline
3Fib & 7 & 2048 
\\
4Fib & 7 & 6250 
\\
\end{tabular}\par}\ \\
\caption{%
 Performance of various algorithms for a target bias of at least $7\eps_0$.
}
\label{tab:biases>=7}
\end{table}

\begin{table}[here]
{\centering
 \begin{tabular}{l|c|c}
&
\#spins &
\parbox{13ex}{Run-time \\ (reset steps)}
\\\hline\hline
$\delta$-Fib & 8  & 40320
\\
PAC2 & 13 & 1457 
\\
PAC3      & 10  & 2378 
\\
PPA      & 6  & 204 
\\\hline
2PAC      & 11 & 4687 
\\
4PAC      & 9 & 8201 
\\
6PAC      & 9 & 33321 
\\\hline
3Fib & 8 & 8192 
\\
4Fib & 8 & 31250
\\
\end{tabular}\par}\ \\
\caption{%
 Performance of various algorithms for a target bias of at least $11\eps_0$.
}
\label{tab:biases>=11}
\end{table}

\begin{table}[here]
{\centering
 \begin{tabular}{l|c|c}
&
\#spins &
\parbox{13ex}{Run-time \\ (reset steps)}
\\\hline\hline
$\delta$-Fib & 9  & 362880 
\\
PAC2 & 15 & 4373 
\\
PAC3      & 11  & 5741 
\\
PPA      & 6  & 529 
\\\hline
2PAC      & 11 & 4687 
\\
4PAC      & 11 & 73811 
\\
6PAC      & 9 & 33321 
\\\hline
3Fib & 9 & 32768 
\\
4Fib & 8 & 31250 
\\
\end{tabular}\par}\ \\
\caption{%
 Performance of different algorithms for a target bias of at least $15\eps_0$.
}
\label{tab:biases>=15}
\end{table}

%
\section{Close-up view of cooling curves for 4PAC, 6PAC and $\infty$PAC}
\label{app:mPAC-zoom}
\begin{figure}
\includegraphics[height=0.8\columnwidth,angle=0]{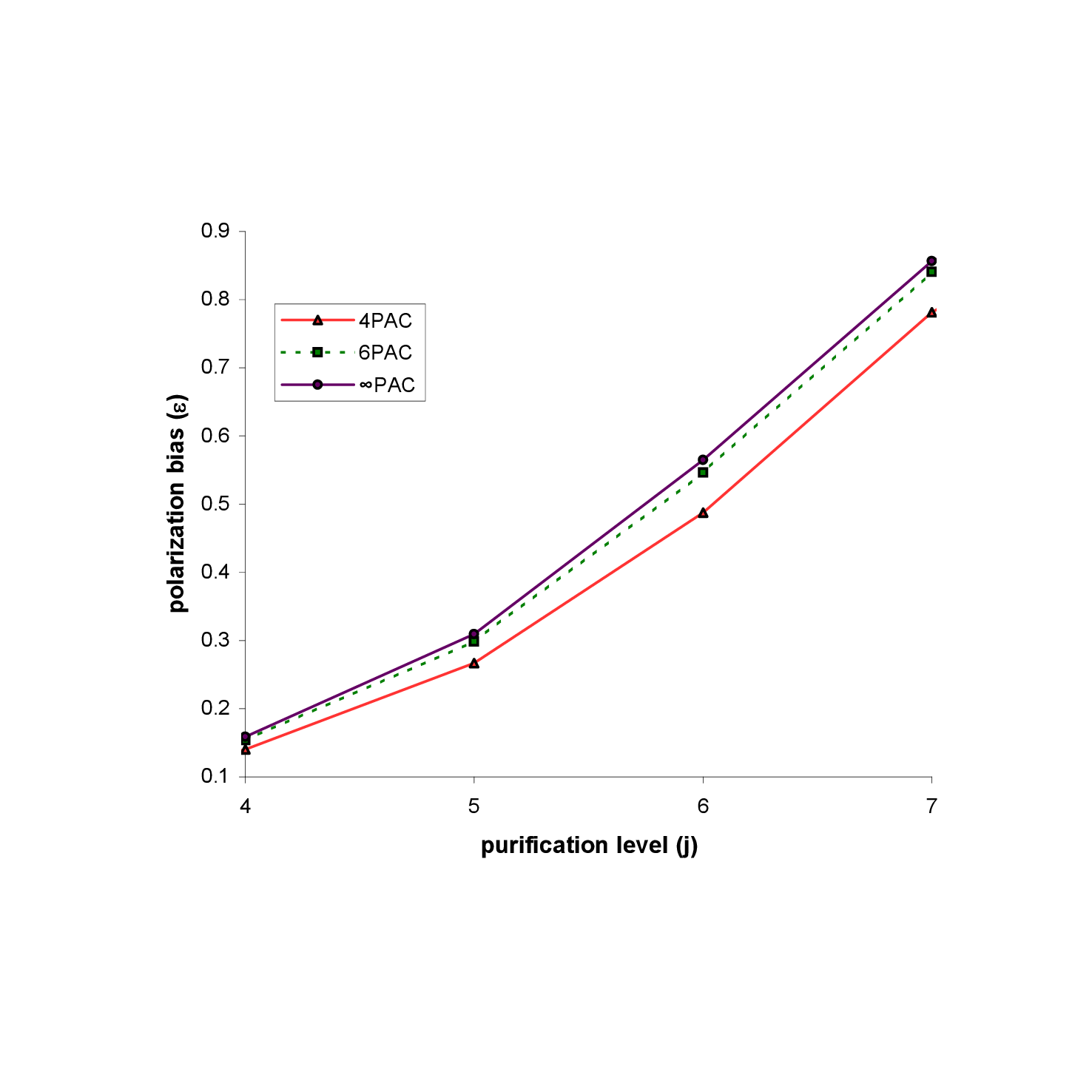}
\caption{Close-up view of the polarization biases ($\eps_{2j+1}$) for 4PAC,
 6PAC and $\infty$PAC as a function of the purification level $j,$ starting
from an initial bias of $\eps_0=0.01$. 
For more details see Figure~\ref{fig:mPAC-analysis-for-0.01}.}
\label{fig:mPAC-zoom-for-0.01} 
\end{figure}
\begin{figure}
\includegraphics[height=0.8\columnwidth,angle=0]{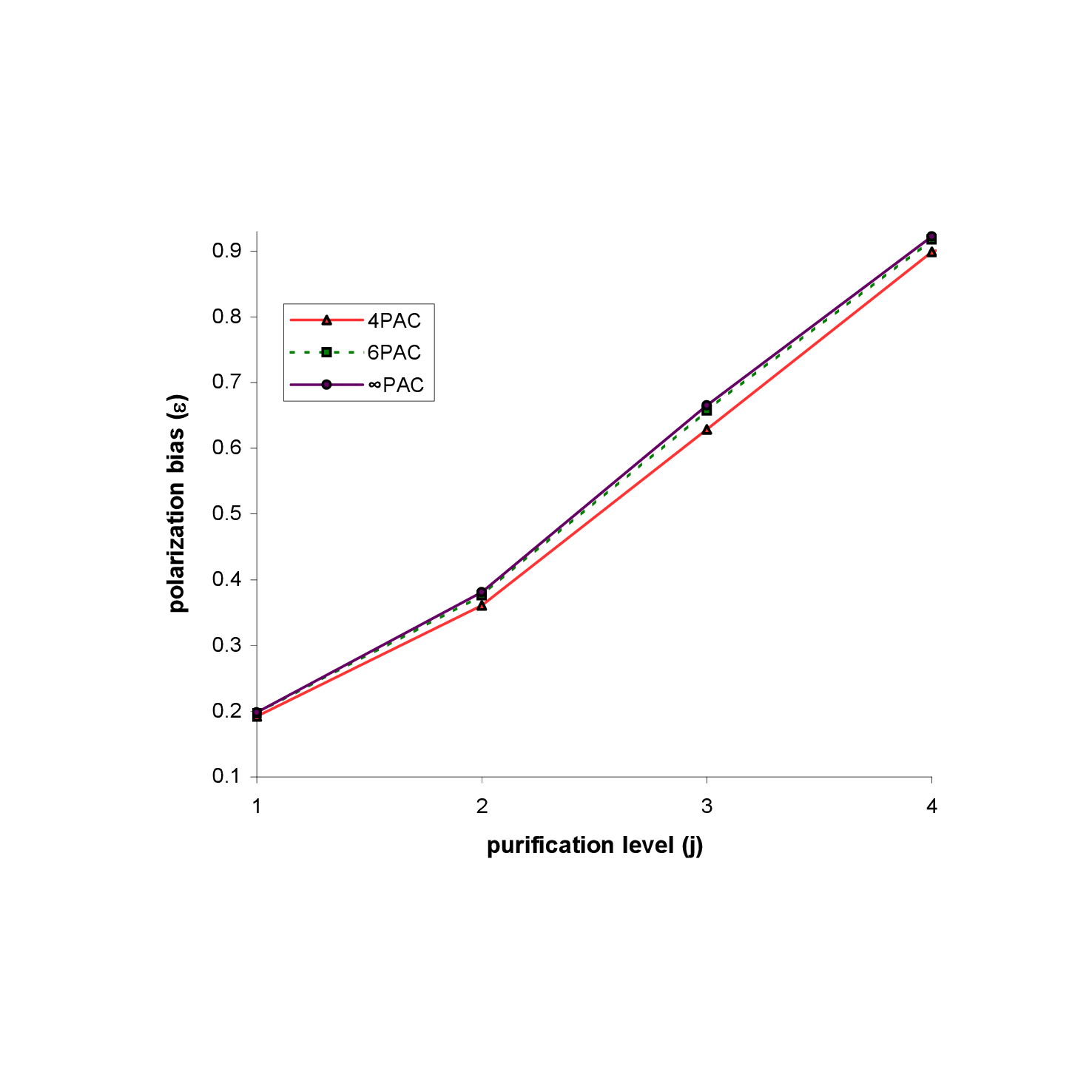}
\caption{Close-up view of the polarization biases ($\eps_{2j+1}$) for 4PAC,
 6PAC and $\infty$PAC as a function of the purification level $j,$ starting
from an initial bias of $\eps_0=0.1$. 
For more details see Figure~\ref{fig:mPAC-analysis-for-0.1}.}
\label{fig:mPAC-zoom-for-0.1} 
\end{figure}

\end{document}